\documentclass[12pt,preprint]{aastex}

\slugcomment{Submitted: ApJ, April 4, 2010; Revised May 11, 2010}

\shorttitle{Decomposition of Cepheid Light Curves}
\shortauthors{Freedman \& Madore}

\begin{document}



\title{\rm \bf A Physically-Based Method for Scaling
  Cepheid Light Curves for Future Distance Determinations}


\author{\bf Wendy L. Freedman \& Barry F. Madore}\affil{The Observatories \\ Carnegie
Institution for Science \\ 813 Santa Barbara St. \\ Pasadena, CA
~~91101} \email{wendy@obs.carnegiescience.edu, barry@obs.carnegiescience.edu}



\begin{abstract}
We present a technique for decomposing Cepheid light curves into their
fundamental constituent parts; that is, their radius and temperature
variations. We demonstrate that any given pair of
optical luminosity and color curves can  be used to predict the
shape, amplitude and phase of a Cepheid's light variation at any
other wavelength.  With such predictions in hand, a single new
observation at any given new wavelength can be used to normalize the
properties of the predicted light curve, and in specific, 
derive a precise value of the time-averaged mean. We
suggest that this method will be of great advantage in efficiently
observing and precisely obtaining the mean properties of known
Cepheids scheduled to be observed at new wavelengths, specifically in
the mid-infrared where JWST will be operating.
\medskip
\medskip
\medskip
\medskip\medskip
\medskip
\medskip\medskip
\medskip
\medskip\medskip
\medskip
\medskip\medskip
\medskip
\medskip
\medskip
\medskip
\medskip
\medskip
\medskip
\medskip
\medskip
\medskip
\medskip
\medskip
\medskip
\medskip
\medskip
\medskip
\medskip
\medskip
\medskip
\medskip
\medskip
\medskip
\medskip
\medskip
\medskip
\medskip
\medskip
\medskip
\medskip
\medskip
\medskip
\medskip
\medskip
\medskip
\medskip

\end{abstract}

{\it ``The variable star $\delta$ Cephei, the prototype of its class, has
been observed so long and so often -- visually, photographically, and
photoelectrically -- that little new can be expected from additional
observations, no matter what their quality.''} -- Stebbins (1945)
\medskip
\section{Introduction}

The absolute luminosities of stars can be simply expressed as the
product of only two quantities: the total area of the radiating
surface and the mean surface brightness of that same area. For
spherical stars the area is readily calculated from the radius; and
the surface brightness is locally controlled by the temperature. For a
radially pulsating star, such as a Cepheid, its time variation in
luminosity is also controlled by the coupled time variation of those
same two physical parameters, radius and temperature.  In this paper,
we explore this property further, enabling us to use data obtained at
multiple wavelengths to provide constraints on these parameters.

The basic relations are captured by Stefan's Law in the definition of
the effective temperature ($T_{e}$) and its derivative. Bolometrically

$$ L = 4\pi R^2 \sigma T_{e}^4 $$

\par\noindent
and 

$$  {\Delta L}/{L} = {2 \Delta R}/{R} + {4\Delta T_{e}}/{T_{e}} + constant $$

\par\noindent The differential form clearly demonstrates that the
luminosity variation should be readily decomposed into two terms, a
radius and a temperature term. Since the temperature of a star is
readily estimated from its (optical) colors, it follows that the
surface area variation must be whatever residual remains after the
color variation is scaled (to a surface brightness) and subtracted
from the (composite) luminosity curve. This suggests a path whereby
the two physical terms can be decoupled empirically and evaluated
independently.

\begin{figure}
\includegraphics [width=15cm, angle=270] {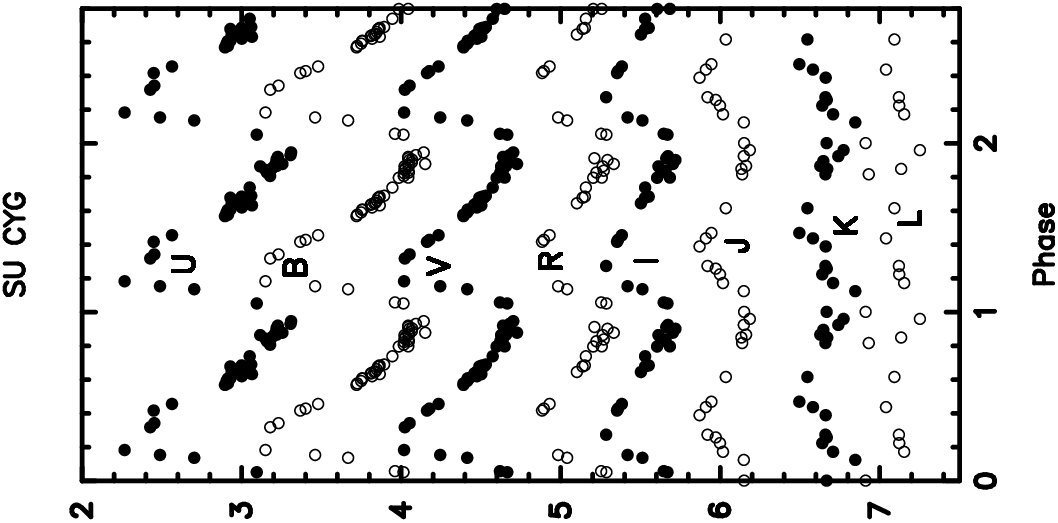}
\includegraphics [width=15cm, angle=270] {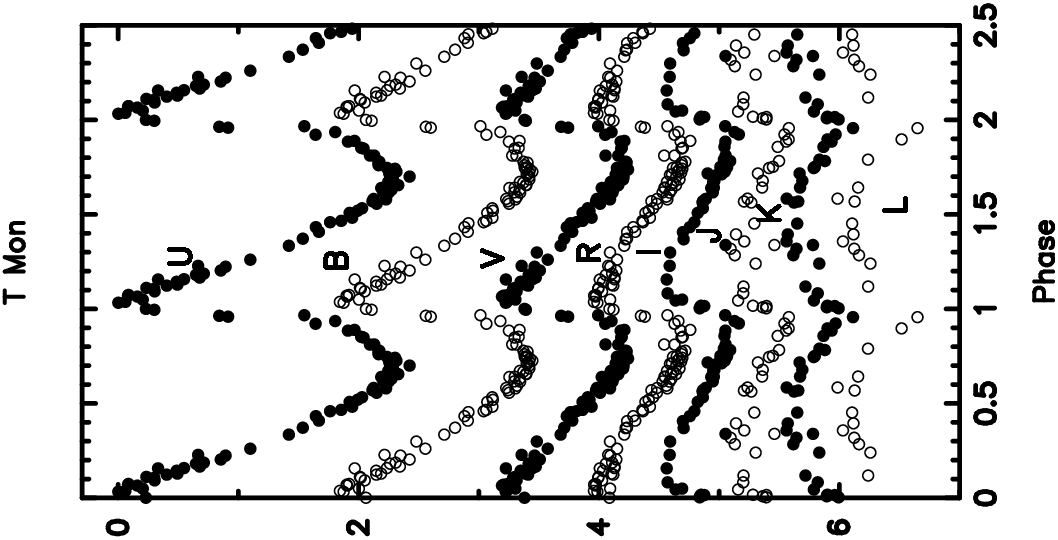}
\caption{Multi-wavelength light curves for two
Galactic Cepheids, SU~Cyg and T~Mon. The run of
amplitude and phase of maximum light for Cepheid light curves as a
function of wavelength of the observations is easily seen in these
data drawn primarily from Wisniewski and Johnson (1968).  Moving from
top to bottom the amplitudes decrease as the wavelength increases. The
phase of maximum light at the progressively longer wavelengths falls
systematically behind the timing of maximum light at the shorter
wavelengths.  Finally, it is quite apparent that the overall shape of
the light curve also changes with wavelength. In the blue there is a
steep rise followed by a slower, linear decline. In the near infrared
the light curve is more symmetric and  cycloidal in nature,
showing a wide maximum flanked symmetrically by rapid dips to
short-lived minima.}
\end{figure}

The above equations apply to the bolometric luminosity. Does this
simplicity transfer over to empirically measured broad-band photometry? At
first sight, multi-wavelength observations (of Cepheids in particular)
might suggest that the situation is far more complicated, given that
Cepheid light curves at different wavelengths have fundamentally
different shapes, systematically changing amplitudes and advancing
phases of maximum light as a function of wavelength. See Figure 1 for
examples.


Early observers of Cepheids noted that the amplitudes of those
variable stars systematically decreased as a function of advancing
wavelength (Wisniewski \& Johnson 1968), where it was noted that the shapes of Cepheid light
curves became less triangular and more cycloidal in form as their
amplitudes dropped. It was also seen that the timing of maximum light
for Cepheids was shifting to later phases, again as longer and longer
wavelengths were explored (see for example the atlas of
multi-wavelength light curves in Wisniewski \& Johnson).
Stebbins (1945) is the first to have noted the decreasing amplitude a
Cepheid (by a factor of 3.4 for his observations of $\delta$ Cep) with
increasing wavelength of his observations (from U[3530\AA] to
I$_S$[10,300\AA]), and the concommitent shifting of maximum light to
later phases (by +0.05 in phase over the same, U to I, wavelength
range).  The first practical use of these trends was made by Freedman
(1988; see below).

The photometric behavior described above can be understood as follows:
The asympototic (long-wavelength) behavior of a Cepheid light curve is
dominated by the (geometric) radius variation. It is well known from
integrating radial-velocity curves of Cepheids that the form of the
radius variation with time is well described as being cycloidal in
shape (see, for example, radius and radial velocity data presented by
Imbert 1981). Since it is geometric in nature, the contribution of the
radius variation to the light curve of a Cepheid will be largely
independent of wavelength, modulo small (wavelength-dependent) optical
depth effects. If there were no additive temperature variations, any
given Cepheid's light curve would have the same shape at all
wavelengths; but observations indicate otherwise. A test can be made
by subtracting two light curves obtained at different wavelengths. The
differencing directly cancels out the equally-contributing radius
term, and then leaves a pure temperture-driven term. At short
wavelengths color is primarily driven by temperature. Clearly then, by
appropriately scaling the color curve and subtracting it (in magnitude
space) from the luminosity curve one will be left with the areal
variation (i.e., radius) curve. These same precepts are at the heart
of the Baade-Wesselink method, where photometry and radial velocities
are combined to determine absolute radii and metric distances. The
method we are describing here provides a way of measuring precision
distances without the need for  radial velocity measurements.

\begin{figure}
\includegraphics [width=12cm, angle=270] {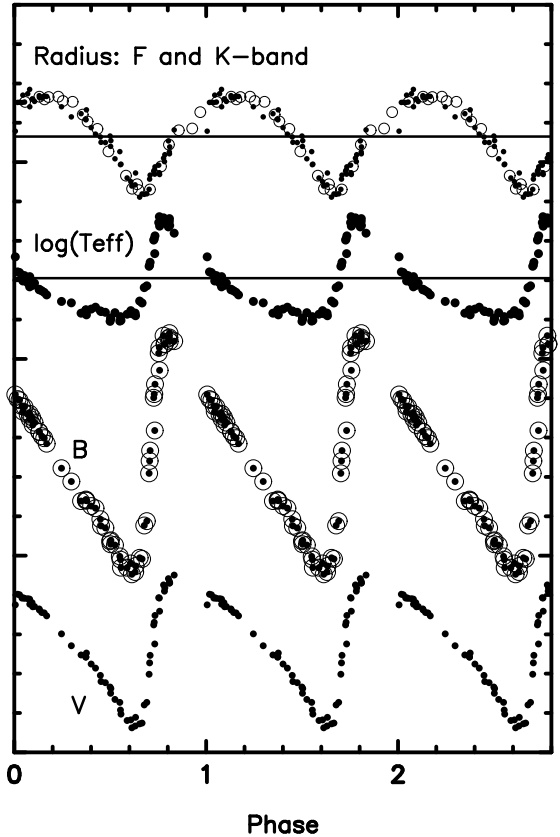}
\caption{An example of the physical decomposition of a Cepheid light
curve. The B and V light curves for the Large Magellanic Cloud
Cepheid, HV~2257, are shown in the lower part of the plot. They act as
the generating functions for the two curves plotted at the top of the
figure.  The larger filled circles, marked log~T$_{eff}$, define 
the (B-V) color curve. The upper plot is a composite of K-band,
near-infrared data (K: open circles), and the CPR predicted
radial-displacement curve (F: filled circles) obtained by scaling and
subtracting the temperature variation
from the observed luminosity curves leaving the radius variation. The
similarity in the K and F curves is striking.}
\end{figure}

\section{Extracting Radius Curves from Optical Light Curves}

In a study of the Local Group dwarf galaxy, IC~1613, Freedman (1988)
showed that one could use well-sampled B-band light curves, for
example, to predict the amplitudes at I-band wavelengths, adjust the
phasing appropriately, and predict the mean magnitude at I, based on a
single observation at that longer wavelength. The scaling relations,
rephasing and predicted light-curve shapes were only approximate, but
given the factor of 2-3$\times$ systematic reduction in amplitude in
going from B to I the method provided a significant improvement to the
mean magnitudes.

Subsequently, a number of follow-up studies introduced greater degrees
of mathematical sophistication, a larger variety of templates (Stetson
1996; Labhardt, Sandage \& Tammann 1997) and more complicated fitting
procedures, based, for example, on Fourier decomposition (e.g., Ngeow
et al. 2003; Soszynski, Gieren \& Pietrzynski 2005), principal
component analysis (as in Tanvir et al. 2005; Kanbur \& Mariani 2004),
or a combination of both (Yoachim et al. 2009).  Labhardt, Sandage \&
Tammann concluded that except for a brief period of time, during the
ascending phase of the light curve, their method and that of Freedman
(1988) had the same precision (see their Figure 2); while Fourier
decomposition methods suffer from global fitting instabilities when
samples are sparse or highly clumped, and especially when higher-order
terms are included. Nevertheless, with care each of the methods is
capable of scaling light curves across wavelengths and reducing the
scatter in the predicted/fitted mean.  All of these methods are
phenomenological.  In the following we describe a physically-motivated
approach, which is simpler and more precise. We refer to this as
the Carnegie Photometric Radial-Displacement (CPR) Method.

\section{The Carnegie Photometric Radial-Displacement (CPR) Method}

Scaling and subtracting the temperature variation to reveal the pure
radius variations was first discussed by Madore (1985) while
attempting to produce a function that was minimally impacted by phase.
The so-called Feinheit function combines the luminosity and color in
such a way so as to minimize the amplitude of the resulting light
variation. This function, F = V - $\alpha$$\times$(B-V), reaches a
minimum amplitude when the temperature variation is cancelled, leaving
only the radius variation contributing to the light variation.  As it
turns out the color coefficient ``$\alpha$'' in the definition of the
Feinheit function is directly related to ``a'' in the 
Baade-Wesselink formalism used in calculating the logarithmic surface
brightness from colors (for example, see Kervella et al. 2004 for a recent
application and compilation). The Feinheit function then is
numerically equivalent to the so-called ``photometric radial
displacement''. The values of the Feinheit color coefficients
``$\alpha$'' statistically correlate with period (see Figure 3 in
Madore 1985). It is reasonably expected that they are physically
controlled by temperature (i.e., more directly correlated with
intrinsic color.)


\begin{figure}
\includegraphics [width=12cm, angle=270] {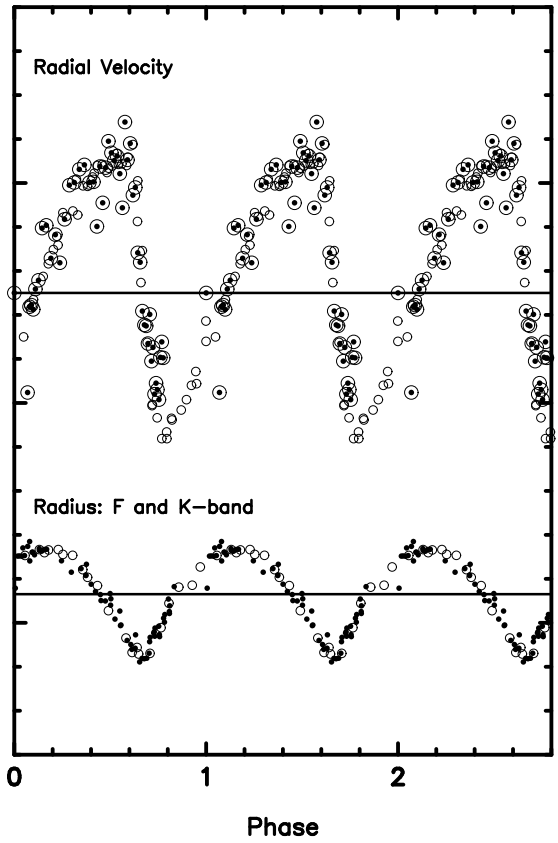} 
\caption{A demonstration that F and K both represent
radius variations. Lower plot: The CPR predictions (F: filled circles)
overplotted on K-band light curve (K: open circles) for the LMC
Cepheid HV~2257, as in Fig. 2. The upper plot shows the published
radial velocity curve (open circles, from Imbert et al. 1985; Imbert
1987) with the scaled first derivatives of the CPR data overplotted
(circled dots).}
\end{figure}

In Figure 2 we first demonstrate the physical decomposition of an
optical light curve into its temperature and radius components.  We
begin with the B and V light curves as plotted in the lower portion of
the figure. We note the larger amplitude of the B-band light curve as
compared to V. It is also easily seen, in comparison, that the B-band
light curve has much more linear decline than the more rounded
descending portion of the V-band light curve. A straightforward,
point-by-point differencing of the two light curves results in the
color curve plotted just above the B-band light curve. Given that
color maps directly to (logarithmic) temperature we have labelled that
curve as log(T$_{eff}$). We next scale the log(T$_{eff}$) curve and
subtract its contribution from the B-band light curve giving rise to
the radius curve delineated by the filled circles plotted across the
top of the panel. The K-band light curve is widely considered to be
dominated by pure radial variations;  those data are plotted as open
circles in the same portion of the diagram. The coincidence is
impressive given that only a vertical shift in the two magnitude
scales was allowed.

In Figure 3 we make our final argument for F being identified with the
radial displacement. The predicted radius curve (F: filled circles)
and the overplotted K-band data (K: open circles) as already given in
Figure 2 (above) are reproduced, for convenience, at the bottom of this
figure. We then take the first derivative of the K-band light curve
and plot it as circled dots in the upper portion of the frame (labeled
``Radial Velocity''). Over-plotted in that same part of the frame are
published radial velocities for HV 2257 from (Imbert et al. 1985;
Imbert 1987). The only adjustment was the amplitude of the variation
to match the logarithmic scaling of the photometric data. Below, we
quantify the level of this agreement. We conclude that K, F and radius are
equivalent representations that allow a successful decomposition of
optical light curves into physical quantitites.

However, before moving on we offer the following discussion prompted by 
the kind suggestions made by the referee. Consider the following
equations describing the decomposition of the light variation in two
distinct bands (B and V, for example) due to areal variations
parameterized by $\theta$ and variations in the surface brightness
induced by temperature variations, $T_{eff}$

$B = B_0 - 5~ log(\theta) - C_B \times 2.5~log(T_{eff})$

$V = V_0 - 5~ log(\theta) - C_V \times 2.5~log(T_{eff})$

\noindent 
Recalling that F = V - $\alpha$(B-V) this translates into 

$F = [V_0 - \alpha(B_0 - V_0)] - 5~ log(\theta) - [C_V - \alpha(C_B - C_V)] \times  2.5~log(T_{eff})$

So by setting $\alpha$ = $C_V/(C_V - C_B)$ the Feinheit function (by
design) eliminates the temperature variation.  If the
color-temperature relation is non-linear then the values of $C_B$ and
$C_V$ will themselves change with temperature and so $\alpha$ will
also be expected to be a function of intrinsic color (i.e.,
temperature) also.  Since longer-period Cepheids are statistically
redder (cooler) that short-period Cepheids it follows that $\alpha$
will also be a (statistical) function of period, which explains the
correlation that was found and reported in Madore (1985), as stated
earlier.

We now explore how the CPR method can be used in a practical sense
through a series of decomposition and reconstruction examples. These
range from Galactic examples with dozens of observations covering half
a dozen or more wavelengths, to LMC Cepheids with extensive optical
and near-infrared data, to Cepheids at the edge of detectability in
more distant galaxies where  one or two observations are available
beyond their (optical) discovery wavelengths.

We now consider a Galactic Cepheid, YZ Cyg, for a further
demonstration of the CPR decomposition and prediction method. This
time we use BV data to predict the details of the slightly
longer-wavelength light curve in the R band, but this time also to
predict the light curve behavior at the shorter U-band wavelength.  In
the lower panel the published UBVR observations are shown as open
circles.  At the top of the panel are smoothed versions of the radius
variation (as derived only from the BV light curves), and the
temperature curve again derived exclusively from the BV light
curves. Now immediately above the R-band observations is the predicted
(solid line) R-band light curve based on the reprojected radius and
temperature curves.  Below the U-band data is the (solid line)
prediction for the ultraviolet light curve. To the eye they are
indistinguishible. We now quantify that match.

\begin{figure}
\includegraphics [width=13cm, angle=270] {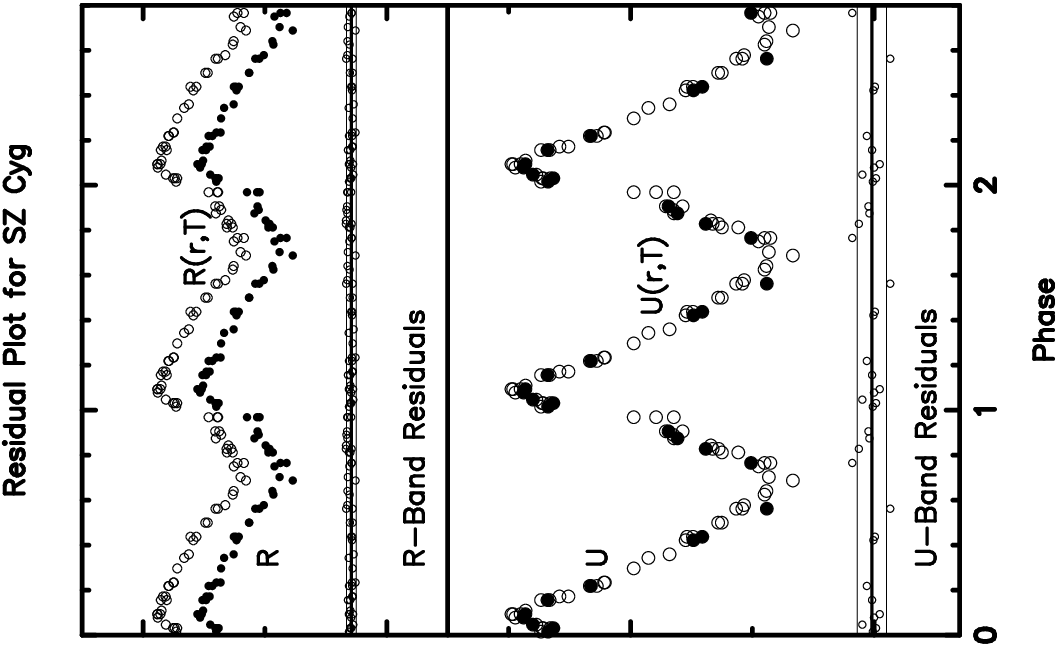} 
\includegraphics [width=13cm, angle=270] {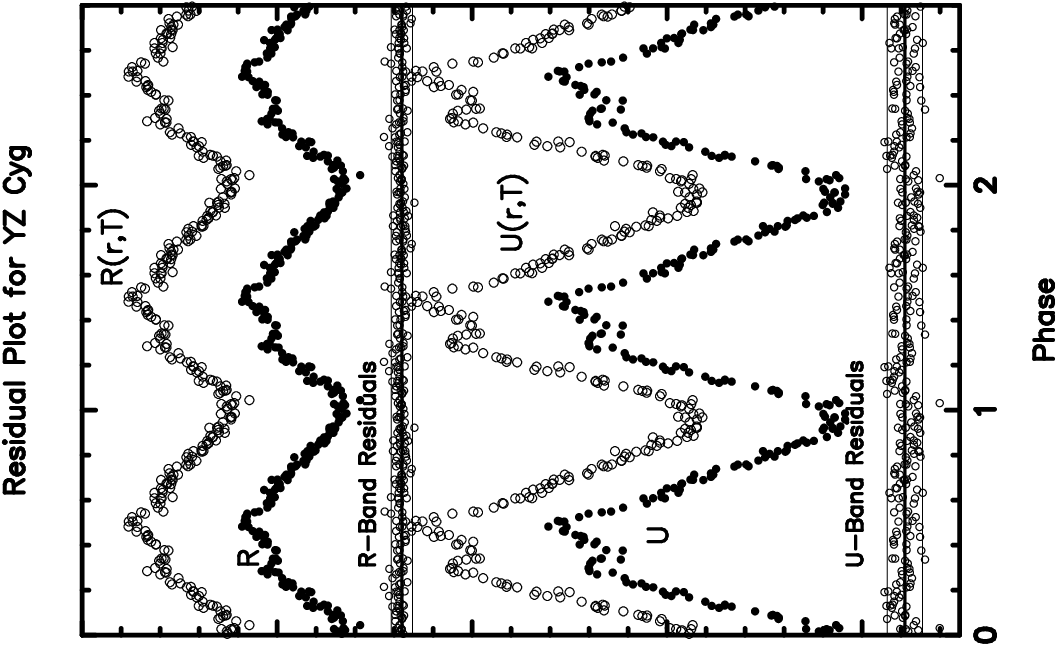}
\caption{Point-by-point CPR reconstructions of the R (upper) and
U (lower) light curves based upon predictions from the B and V data
alone. The two flat-lined scatter plots below each pair of light
curves are the individual differences between the observed and the
predicted light curves. No significant change in scatter nor any
systematic deviations from zero is seen as a function of phase.
To within the scatter of the observations, no further parameter
characterizing the light curves is indicated; in fact, the fits are
consistent with being dispersionless.}
\end{figure}

In Figure 4 we isolate the R \& U-band data and their predictions now
showing the point-by-point reprojection of the combined radius and
temperature curves back into the observational planes.  Below each
curve is the run of differences between the observed and the predicted
light curves. Because no smoothing in the physical plane was applied,
the residuals carry the full complement of the uncorrelated errors
associated with all three of the B, V and R (B, V and U) data points
that went into the various solutions. The calculated scatter in R-band
comparison is $\pm$0.013~ mag; the scatter in U is $\pm$0.028
mag. The $\pm$2-sigma bounds are plotted around the measured
differences. There are no significant trends of the scatter with phase. The
scatter is consistent with photometric errors in the combined data.

\begin{figure}
\includegraphics [width=12cm, angle=270] {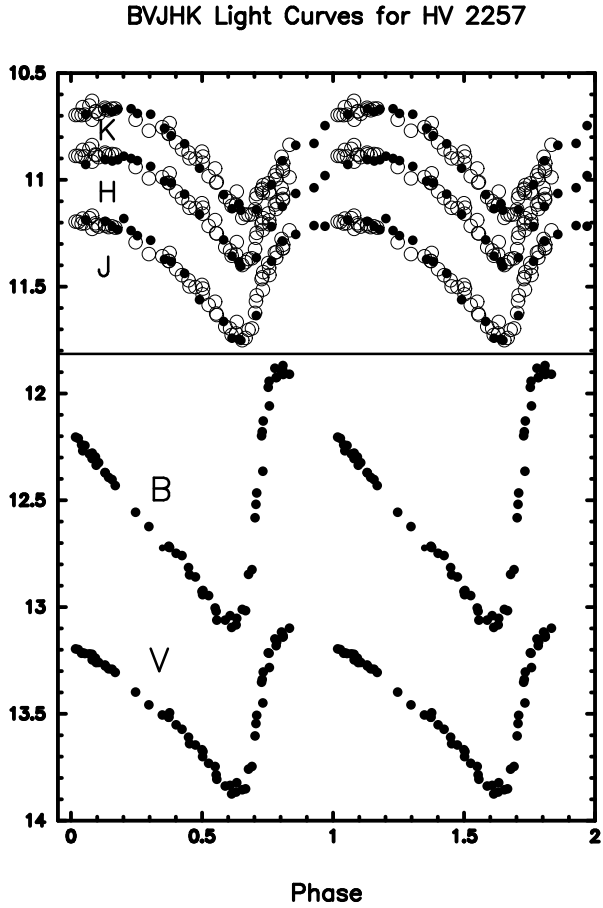}
\caption{Prediction of JHK near-infrared light curves for HV 2257 from
BV optical light curves. Open circles in the upper three plots are CPR
predictions for the amplitude, shape and phasing of the three
near-infrared light curves. The observed data are shown as filled
circles. Excellent agreement is seen. }
\end{figure}

In Figure 5 we show the point-by-point prediction of JH \& K light
curves from the precision BV CCD data for HV 2257 taken from Mofett et
al. (1998). In the lower part of the panel we show the individually
phased BV observations folded over 2.8 cycles. Using these
observations to produce radius and temperature curves we then
reproject these data back into the observational plane but now at the
three near-infrared JHK wavelength bands. Those predictions are shown
by open circles in the upper part of the panel. Overplotted on the
predicted values are the observed JHK data (again as filled circles)
from Persson et al. (2004). Figure 6 shows the results of predicting
lightcurves at longer (R-band) and shorter (U-band) wavelengths than
the basis light curves. To within the observational scatter the
observations and the predictions are indistinguishible.

\begin{figure}
\includegraphics [width=12cm, angle=270] {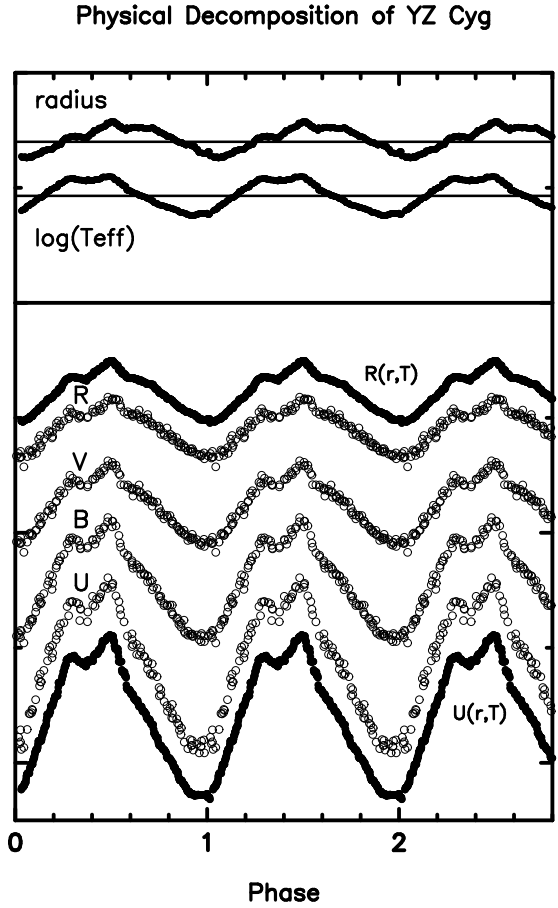}
\caption{UBVR light curves for the Galactic Cepheid YZ~Cyg. Open
circles are the published data from (Wisniewski \& Johnson 1968). At
the top of the panel are the smoothed versions of the radius and
temperature curves made as intermediary functions in the CPR
reconstruction. The solid curves above and below the published-data R
\& U light curves are the CPR predicted light curves, labeled R(r,T)
and U(r,T), respectively. Because of the precision of the predictions
the two curves (predicted and observed) are displaced so that the data
points and the predictions can be seen separately; if over-plotted the
points would be indistinguishible.}
\end{figure}

\begin{figure}
\includegraphics [width=12cm, angle=270] {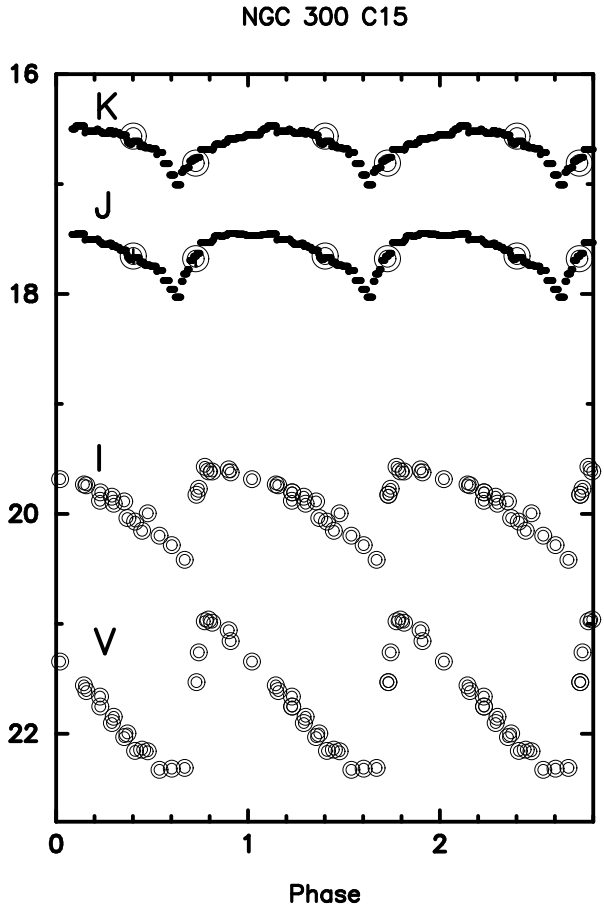} 
\caption{V and I light curves (lower half of the figure) for the
32-day extragalactic Cepheid C15 in NGC~300. The upper panel shows the
CPR predictions based on V and I data (solid dots)  for two
near-infrared bandpasses (K \& J) overlaid upon the only two
random-phase observations (circled circles) available for this
Cepheid. The size of the inner circle closely approximates the quoted
errors on the near-infrared photometry. These data are well fit by the
CPR predictions.}
\end{figure}

We close this series of demonstration cases with an example taken from
the recent literature on multi-wavelength observations of Cepheids
significantly more distant and fainter than our Galactic examples and
about 40 times further away than the LMC. This is the Cepheid C15 in
the South Polar Group spiral galaxy NGC~300. Plotted as circled
circles in the lower half of Figure 7 are V and I data from Gieren
et al.  (2004). From these two wavelengths we constructed temperature
and radius curves, applied the CPR Method and reprojected them into
the K and J band passes for which there are random phase data from
Gieren et al. (2005).  The fit of the predictions to the two
observations in each of J and K are shown at the top of the
figure. Despite having switched from BV to VI for our physical
decomposition basis the method demonstrably works here too in very
accurately phasing up with the observed data.

\begin{figure}
\includegraphics [width=15cm, angle=270] {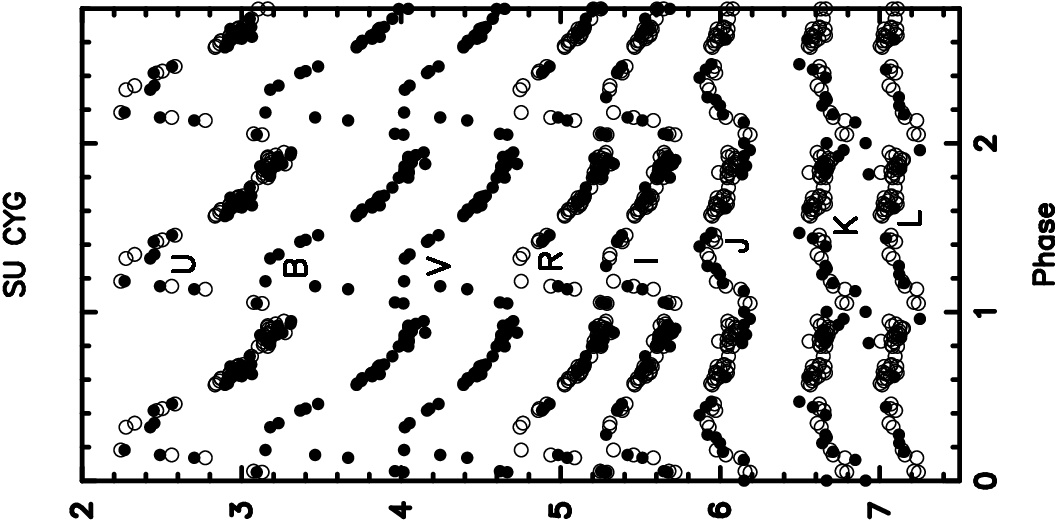} 
\includegraphics [width=15cm, angle=270] {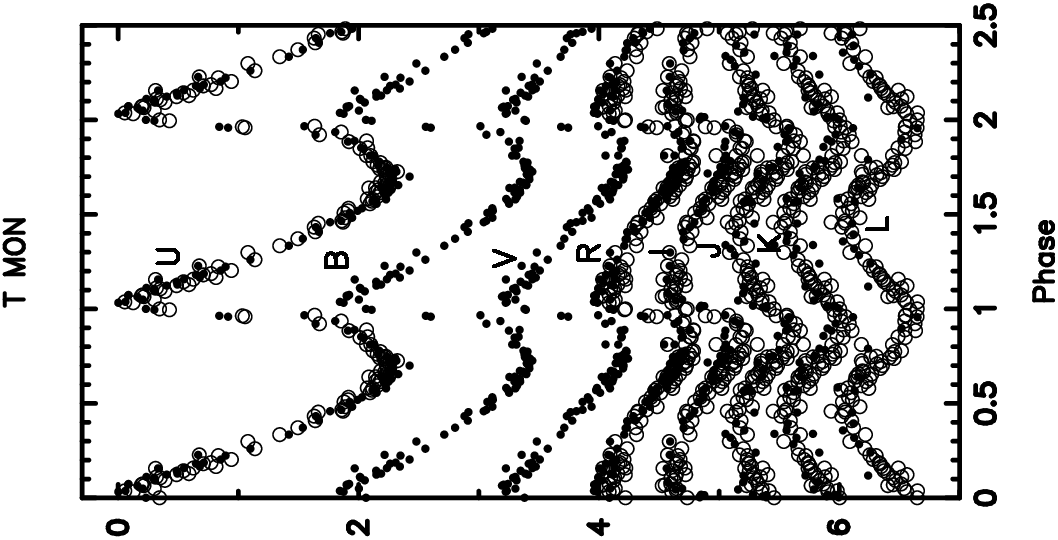} 
\caption{The two Cepheids, SU~Cyg and T~Mon, shown at the beginning of
the article but now with their multi-wavelength light curves
delineated by the original observations (filled circles) overlaid with
CPR predictions (open circles) derived from the BV light curves
alone. The noise in the predictions is directly ascribable to the
photometric noise in the basis-generating B and V light curves.}
\end{figure}

\section{Conclusions}

We have shown that the diversity of Cepheid lightcurve amplitudes,
shapes and relative phases, encountered over all currently available
wavelengths, are reducible to two simple but also physically
meaningful quantitities: radius and temperature.  Given a set of
applicable scaling relations, the radius and temperature curves can be
used to predict the exact form of any given Cepheid's behavior at all
currently observed wavelengths, from the ultraviolet to the
mid-infrared (see Figure 8). Given this predictive tool set one can
now realize enormous observational gains with no loss in
signal-to-noise in extracting mean properties of Cepheids at other
wavelengths.  This is especially true for deriving time-averaged mean
magnitudes and mean colors, derived from sparsely-sampled,
random-phase observations at even longer or shorter wavelengths.

The next paper in this series will present and discuss the exact
calibration of this method, for future applications especially at
mid-infrared wavelengths where JWST will be fully operational.

\section{Acknowledgments}
We than the referee for his/her detailed suggestion to make the
physical-variable decomposition of the Feinheit formalism more
explicit.

\vfill\eject
\noindent
\centerline{\bf References \rm}
\vskip 0.1cm
\vskip 0.1cm

\par\noindent
Freedman, W.~L. 1988, \apj, 326, 691

\par\noindent
Gieren, W. et al. 2004, \aj, 128, 1167

\par\noindent
Gieren, W. et al. 2005, \apj, 628, 695

\par\noindent
Imbert, M. 1981, \aaps, 44, 319

\par\noindent
Imbert, M. 1987, \aap, 175, 30

\par\noindent
Imbert, M. et al. 1985, \aaps, 61, 259

\par\noindent
Kanbur, S.M., \& Mariani, H. 2004, \mnras, 355, 1361

\par\noindent
Kervella, P., Bersier, D., Mourard, D., Nardetto, N., Fouque, P., \& Coude du Foresto, V. 2004, \aap, 428, 587

\par\noindent
Labhardt, L., Sandage, A., \& Tammann, G. A.
1997, \aap, 322, 751, 

\par\noindent
Madore, B.~F. 1985, \apj, 298, 340

\par\noindent
Moffett, T.J., Giren, W.P., Barnes, T.G., \& Gomez, M. 1998, \apjs, 117, 135

\par\noindent
Persson, S.E., Madore, B.F., Kzreminski, W., Freedman, W.L., Roth, M., \& Murphy, D.C. 2004, \aj, 128, 2239

\par\noindent
Soszynski, I., Gieren, W., \& Pietrzynski, G. 2005, \pasp, 117, 823

\par\noindent
Stebbins, J. 1945, \apj, 101, 47

\par\noindent
Tanvir, N.R., Hendry, M.A., Watkins, A., Kanbur, S. M., Berdnikov, L. N., \& Ngeow, C. C.
2005, \mnras, 363, 749

\par\noindent
Wisniewski, W.Z., \& Johnson, H.L. 1968, Comm. Lunar Planet. Lab., No. 112, 57

\par\noindent
Yoachim, P., McCommas, L.P., Dalcanton, J.J., \& Williams, B.F. 2009,  \aj, 137, 4697

\vfill\eject
\end{document}